\documentclass[12pt]{elsart}
\usepackage{amssymb}
\usepackage{amsmath}
\usepackage{latexsym}
\usepackage{graphicx}

\newcommand\R{\mathbb{R}}
\newcommand\N{\mathbb{N}}
\newcommand\I{\mathbb{I}}
\newcommand\Z{\mathbb{Z}}
\newcommand\C{\mathbb{C}}

\newcommand\rmi{\mathrm{i}}
\newcommand\rme{\mathrm{e}}
\newcommand\rmd{\mathrm{d}}

\newcommand{\bei}{\begin{itemize}}
\newcommand{\eni}{\end{itemize}}

\newcommand\deq{\stackrel{\mathrm{def}}{=}}
\def\12{\frac{1}{2}}
\def\e2{\frac{\epsilon}{2}}
\def\vth{\vartheta}

\begin{document}

\begin{frontmatter}







\title{Infinite quantum well: a coherent state approach}

\author{P. L. Garc\'{\i}a de Le\'on$^{1,2}$, J. P. Gazeau$^{1,3}$ and J. Queva$^{1,3}$}
\address{$^1$ Laboratoire Astroparticules et Cosmologie,
B\^{a}timent Condorcet,
10, rue Alice Domon et L\'eonie Duquet,
75205 Paris Cedex 13}


\address{$^2$ Universit\'e Paris Est - Institut Gaspard Monge (IGM-LabInfo), 5 Bd. Descartes, Champs-sur-Marne,
77454 Marne-la-Vall\'ee Cedex 2}

\address{$^3$ Universit\'{e} Paris Diderot-Paris7, B\^{a}timent des Grands Moulins, 75205 Paris Cedex 13}


\ead{pgarcia@apc.univ-paris7.fr, gazeau@apc.univ-paris7.fr, queva@apc.univ-paris7.fr}
\begin{abstract}
A new family of 2-component vector-valued coherent states for the quantum particle motion in an infinite square
well potential is presented. They allow a consistent quantization of the
classical phase space and observables for a particle in this potential. We then study the resulting position and (well-defined) momentum operators. We also consider their mean values in coherent states
 and their quantum dispersions.
\end{abstract}

\end{frontmatter}

\section{Introduction}

Even though the quantum dynamics in an infinite square well potential represents a rather unphysical limit situation,
 it is a familiar textbook problem and a simple tractable model for the confinement of
a quantum particle. On the other hand this model has a serious drawback  when it is analyzed in more detail.
Namely, when one proceeds to a canonical standard quantization,
the definition of a momentum operator with the usual form $-\rmi\hbar \rmd/\rmd x$ has a doubtful meaning. This subject has been
discussed in many places (see \cite{bfv} for instance), and the attempts of circumventing this anomaly range from
self-adjoint extensions \cite{bfv} to $\mathcal{PT}$ symmetry approaches \cite{znojil}.

First of all, the canonical quantization assumes the existence of a momentum operator (essentially) self-adjoint
in $\mathrm{L}^2(\mathbb{R})$ that respects some boundary conditions on the boundaries of the well. As has been shown,
these conditions cannot be fullfilled by the usual derivative form of the
momentum without the consequence of losing self-adjointness. Moreover there exists an uncountable set of
self-adjoint extensions  of such a derivative operator which makes truly delicate
the question of a precise choice based on physical requirements \cite{bfv,vogity}.

When the classical particle is trapped in an infinite well of real interval $\Delta$,  the Hilbert space of quantum
states is $\mathrm{L}^2(\Delta,\rmd x)$ and the quantization problem becomes similar, to a certain extent, to the quantization
of the motion on the circle $S^1$. Notwithstanding the fact that boundary conditions are not periodic but impose instead
that the wave functions in position representation vanish at the boundary, the momentum operator $\widehat{p}\,$
for the motion in the infinite well should be the counterpart of the angular momentum operator $\widehat{L}$ for the
motion on the circle. Since the energy spectrum for the infinite square well is $\{n^2, \, n\in \N^{\ast}\}$,  we
should expect that the spectrum of $\widehat{p}$ should be $\Z^{\ast}$, like the one for $\widehat{L}$ without
the null eigenvalue. This similarity between the two problems will be exploited in the present paper. We will adapt
the coherent states (CS's) on the circle \cite{debgo,kopap,delgo} to the present situation by  constructing two-component vector CS's, in the spirit of \cite{aeg}, as infinite superpositions of spinors eigenvectors of $\widehat{p}\,$.

 In the present note, we  first describe the CS quantization procedure. We  recall the construction of
 the CS's for the motion on the circle and the resulting quantization. We  then revisit  the infinite square
well problem and propose a family  of vector CS's suitable for the quantization of the related
classical phase space. Note that various  constructions of CS's   for the infinite square well have been carried out,
like the one in \cite{jpajpg} or yet the one resting
upon the dynamical $SU(1,1)$ symmetry  \cite{frank}.
Finally, we  present  the consequences of our choice after  examining  basic
quantum observables, derived
from this quantization scheme, like position, energy, and a quantum version of the problematic momentum.
In particular we focus on their
mean values in CS's (``lower symbols'') and  quantum dispersions. As will be shown,
the classical limit is recovered
after choosing  appropriate limit values of some  parameters present in the expression of our  CS's.

\section{\label{sec:level1} The approach via coherent state quantization}

Coherent state  quantization \cite{gahulare,ber,klau2,gapi,gmm,gaga,gahulare1} is an alternative way of representing
classical observables into a quantum system. The states used in it include Glauber and Perelomov CS's
but lie in a wider definition that admits a large range of
state families resolving the identity. Identity resolution is here the crucial condition.

In fact, these states form a frame of reference well suited to  represent classical
quantities and, in that sense, work as a natural quantization procedure which is in one-to-one correspondence
with the choice of the frame. The validity of a precise frame choice is asserted by comparing spectral
characteristics of quantum observables $\widehat{f}$ with data from the observational space. Unlike canonical
quantization where the whole model rests upon a pair of conjugated variables within the Hamilton
formalism \cite{dirac}, here we need the following elements.

First of all let $X=\{x\in X\}$ be a set equipped with a measure $\mu(\rmd x)$,
and let $\mathrm{L}^2(X, \mu)$ be the Hilbert space of square integrable functions $f(x)$ on  $X$:
\begin{equation}
  | f |^2 =  \int_{X} | f(x)|^2 \, \mu(\rmd x) < \infty\, ,\qquad
\langle f_1 | f_2 \rangle  =  \int_{X} \overline{f_1(x)} f_2(x) \, \mu(\rmd x)\, .
\end{equation}
The set $X$ can be taken as the phase space of a particular problem as will be the case in this paper. Next we need a
finite or infinite orthonormal set $\mathbf{S} = \{ \phi_n(x), n=1,2,\dots\}$,
selected among the elements of $\mathrm{L}^2(X, \mu)$. This set spans, by definition, the separable Hilbert subspace
${\mathcal H}_{\mathbf S}$ and must obey the following condition:
\begin{equation}\label{factor}
 0 < {\mathcal N} (x) \equiv \sum_n | \phi_n (x) |^2 <
\infty \ \mbox{almost everywhere}\, .
\end{equation}
Now let us define the family of \textit{coherent} states $\{ | x \rangle\}_{x\in X}$ \underline{in} $ {\mathcal H}_{\mathbf S}$
through the following linear superposition:
\begin{equation}
 | x\rangle \equiv \frac{1}{\sqrt{{\mathcal N} (x)}} \sum_n \overline{\phi_n (x)} | n\rangle\, ,
\end{equation}
where the states $|n\rangle$ are in one to one correspondence with the functions in the set $\mathbf{S}$. This is an injective map $X \ni x \mapsto | x \rangle \in {\mathcal H}_{\mathbf S}$ (which should be continuous with respect to some minimal topology affected to $X$ for which the latter is locally compact): These coherent states have two main features: they are normalized, $\langle  x | x \rangle = 1 $ and crucially, they resolve the identity in ${\mathcal H}_{\mathbf S}$
\begin{equation}\label{resoid}
\int_X | x\rangle \langle x  | \, {\mathcal N}(x)\,\mu(\rmd x)=
\I_{{\mathcal H}_{\mathbf S}}.
\end{equation}
The CS quantization of a {\it classical} observable $f(x)$ on $X$, consists then in associating to $f(x)$ the operator
\begin{equation}
\label{oper}
 \widehat f := \int_X f(x) |x\rangle\langle x| \, {\mathcal N}(x)\,\mu(\rmd x).
\end{equation}
This ``diagonal'' decomposition (in a topological weak sense) may reveal to be valid for a wide class
of operators. The function $f(x) \equiv \widehat {A}_f (x)$ is called upper (or
contravariant) symbol of the operator $ \widehat f$ and is  non-unique in general. On the other hand, the mean value
$\langle x|  \widehat f | x\rangle \equiv \check{A}_f(x)$  is called lower (or covariant) symbol of
$ \widehat f$.

\section{Quantization of the particle motion on the circle $S^1$}

The motion in the infinite square well potential can be seen as a particular case of the motion on the circle
$S^1$, once we have identified the boundaries of the well with each other and imposed Dirichlet conditions on them.
Functions on this domain will behave as pinched waves on a circle so it is useful to expose first the more general case.

Applying our scheme of quantization we can define the CS's on the
circle. The measure space $X$ is the
cylinder $S^1 \times \R = \{ x \equiv (q,p) \, | \, 0 \leq q < 2\pi , \, p,q \in \R \}$,
\emph{i.e.} the phase space of a particle moving on the circle, where $q$  and  $p$ are canonically conjugate variables.
We consistently choose the measure on $X$ as the usual one, invariant (up to a factor) with respect to canonical
transformations: $\mu(\rmd x) = \frac{1}{2\pi}  \, \rmd q\, \rmd p $.
The functions $\phi_n (x)$ forming the orthonormal system needed to construct CS's are suitably weighted
Fourier exponentials:
\begin{equation}
\label{ficir}
\phi_n (x) = \left(\frac{\epsilon}{\pi}\right)^{1/4}\, \rme^{-\frac{\epsilon}{2}(p-n)^2} \, \rme^{ \rmi nq}\, , \qquad n\in \Z \, ,
\end{equation}
where $\epsilon > 0$ can be arbitrarily small. This parameter includes the Planck constant together with the physical
quantities characterizing the classical motion (frequency, mass, etc.). Actually,
it represents a regularization. Notice that the continuous distribution $x \mapsto | \phi_n(x) |^2$ is the
normal law centered at $n$ (for the angular momentum variable $p$). We establish a one-to-one correspondence between
the functions $\phi_n$ and the states $| n\rangle$ which form an
orthonormal basis of some generic separable Hilbert space $\mathcal{H}$ that can be viewed or not as a
subspace of $\mathrm{L}^2(X, \mu(\rmd x))$. CS's, as vectors in  $\mathcal{H}$, read then as
\begin{equation}\label{ccs}
| p, q \rangle =  \frac{1}{\sqrt{{\mathcal N} (p)}}\,  \left(\frac{\epsilon}{\pi}\right)^{1/4} \sum_{n \in \Z}
\rme^{-\frac{\epsilon}{2}(p-n)^2} \, \rme^{- \rmi nq} | n\rangle\, ,
\end{equation}
where the normalization factor
\begin{equation}\label{norci}
\mathcal{N}(x) \equiv \mathcal{N}(p) = \sqrt{\frac{\epsilon}{\pi}}\sum_{n \in \Z}  \rme^{-\epsilon (p-n)^2} < \infty\, ,
\end{equation}
is a periodic train of normalized Gaussian functions and is proportional to an elliptic Theta function. Applying the
Poisson summation yields the alternative form:
\begin{equation}
\label{Poisnorci}
\mathcal{N}(p) = \sum_{n \in \Z}  \rme^{2\pi \rmi np}\,  \rme^{-\frac{\pi^2}{\epsilon} n^2}\, .
\end{equation}
From this formula it is easy to prove that $\lim_{\epsilon \to 0}\mathcal{N}(p) = 1$.

The CS's (\ref{ccs}) have been previously proposed, however through quite different approaches, by  De Bi\`evre-Gonz\'alez (1992-93) \cite{debgo},
Kowalski-Rembieli\'nski-Papaloucas (1996) \cite{kopap}, and Gonz\'alez-Del Olmo (1998)  \cite{delgo}.

\subsection{Quantization of classical observables}

The quantum operator acting on ${\mathcal H}$, associated to the classical observable $f(x)$, is obtained as
in (\ref{oper}). For the most basic one, i.e. the classical observable $p$ itself, the procedure  yields
\begin{equation}\label{psym}
\widehat p  = \int_{X}   \mathcal{N}(p)\, p\, | p,q \rangle \langle p, q | \mu (\rmd x) = \sum_{n \in \Z}
n\, | n\rangle \langle n| ,
\end{equation}
and this is nothing but the angular momentum operator, which reads in angular position representation
(Fourier series): $ \widehat{p} = -\rmi\frac{\partial}{\partial q}$.

For an arbitrary function  $f(q)$, we have
\begin{align}
\widehat{f(q)}& =  \int_{X} \mu (\rmd x)  \mathcal{N}(p) f(q) \,  | p,q \rangle \langle p, q | \nonumber \\
&= \sum_{n,n' \in \Z}
 \rme^{-\frac{\epsilon}{4}\,(n-n')^2} \,c_{n-n'}(f)| n\rangle \langle n' | ,
\label{f(beta)}
\end{align}
 where $c_{n}(f)$ is the $n$-th Fourier coefficient of $f$. In particular, we have for the
angular position operator $\widehat{q}\,$:
\begin{equation}
\label{opangle}
  \widehat{q} = \pi \I_{{\mathcal H}} + \rmi \sum_{n\neq n'}
 \frac{ \rme^{-\frac{\epsilon}{4}(n-n')^2}}{n-n'}\, | n\rangle \langle n' |\, .
 \end{equation}
The shift operator is the quantized counterpart of  the  ``Fourier fundamental harmonic'':
\begin{equation}
\label{opfourier}
 \widehat{ \rme^{\rmi q}} =   \rme^{-\frac{\epsilon}{4}}\, \sum_{n}
 | n + 1\rangle \langle n |.
\end{equation}
The commutation rule between (\ref{psym}) and (\ref{opfourier}) gives
\begin{equation}
[ \, \widehat{p},  \widehat{ \rme^{\rmi q}} \,] =  \widehat{ \rme^{\rmi q}},
\end{equation}
and is canonical in the sense that it is in exact correspondence with the classical Poisson bracket
\begin{equation}
\left\{ p,  \rme^{\rmi q} \right\} = \rmi  \rme^{\rmi q}.
\end{equation}
Some interesting aspects of other such correspondences are found in \cite{rabeie}. For arbitrary functions of $q$
the commutator
\begin{equation}
 [ \, \widehat{p},  \widehat{f(q)} \,] =  \sum_{n, n'}(n-n')
 \rme^{-\frac{\epsilon}{4}(n-n')^2}\,c_{n-n'}(f)\, | n\rangle \langle n' |,
\end{equation}
can arise interpretational difficulties. In particular, when $f(q)=q$, i.e. for the angle operator
\begin{equation}
\label{ccrcir}
 [ \,\widehat{p}, \widehat{q} \,] = \rmi \sum_{n \neq n'}
  \rme^{-\frac{\epsilon}{4}(n-n')^2}\, | n\rangle \langle n' |\, ,
 \end{equation}
the comparison with the classical bracket $\left\{ p, q \right\} = 1$ is not direct.
Actually, these difficulties are only apparent if we consider instead the $2\pi$-periodic extension to $\mathbb{R}$ of  $f(q)$.
The position observable $f(q)=q$, originally defined in the interval $[0,2\pi)$, acquires then a sawtooth
shape and its periodic discontinuities are accountable for the discrepancy. In fact the obstacle is circumvented if
we examine, for instance, the behaviour of the corresponding lower symbols at the limit $\epsilon \to 0$. For the
angle operator we have
\begin{align}
\label{lowsymb}
\nonumber \langle p_0, q_0 | \, \widehat{q} \, |  p_0, q_0 \rangle    &=   \pi + \frac{1}{2}\,
\left(1 + \frac{\mathcal{N}(p_0 - \frac{1}{2})}{\mathcal{N}(p_0)}\right)\, \sum_{n \neq 0} \rmi\,
\frac{ \rme^{-\frac{\epsilon}{2}n^2 + \rmi n q_0}}{n} \\
& \underset{\epsilon \to 0}{\sim}     \pi +  \sum_{n \neq 0} \rmi\, \frac{ \rme^{\rmi n q_0}}{n}\,,
\end{align}
where we recognize at the limit the Fourier series of $f(q)$.
For the commutator, we recover the canonical commutation rule modulo Dirac singularities on the
lattice $ 2\pi \Z$.
\begin{align}
\label{symbcom}
\nonumber   \langle p_0, q_0 |  [\, \widehat{p},  \widehat{q}\, ]\, |  p_0, q_0 \rangle &=  \frac{1}{2}\,\left(1 +
\frac{\mathcal{N}(p_0 - \frac{1}{2})}{\mathcal{N}(p_0)}\right)\left( -\rmi  + \sum_{n\in \Z} \rmi  \rme^{-\frac{\epsilon}{2}n^2
+ \rmi n q_0}\right) \\
& \underset{\epsilon \to 0}{\sim}   -\rmi +  \rmi\sum_{n } \delta(q_0 - 2 \pi n).
\end{align}

\section{Quantization of the motion in an  infinite well potential}

\subsection{The standard quantum context}
Any quantum system trapped inside the infinite square well $0 \leqslant q
\leqslant L$ must have its wave function equal to zero beyond the boundaries. It is thus natural to impose
on the wave functions the conditions
\begin{equation}\psi (q) = 0, \qquad q \geqslant L \quad \mbox{and} \quad q \leqslant 0\, .
\label{3.1}
\end{equation}
Since the motion takes place only inside the interval $[0, L]$, we may
as well ignore the rest of the line and replace the constraints (\ref{3.1}) by the following ones:
\begin{equation}
\label{domH}
\psi \in \mathrm{L}^2([0, L],\rmd q), \qquad \psi (0) = \psi (L) = 0\,.
\end{equation}
Moreover, one may consider the periodized well and instead impose the
cyclic boundary conditions $\psi (n L) = 0, \, \forall n \in \Z$.

In either case, stationary states of the trapped particle of mass $m$ are
easily found from the eigenvalue problem
for the Schr\"odinger operator with Hamiltonian:
\begin{equation}
H \equiv H_{\rm w} = - \frac{\hbar^2}{2m} \frac{\rmd^2}{\rmd x^2} \, .
\label{3.2}
\end{equation}
This Hamiltonian is self-adjoint \cite{simon} on an appropriate dense  domain in (\ref{domH}).
Then
\begin{equation}
\Psi (q,t) =  \rme^{-\frac{\rmi }{\hbar} Ht} \Psi (q,0)\, ,
\label{time-evol}
\end{equation}
where $\Psi (q,0) \equiv \psi (q)$ obeys the eigenvalue equation
\begin{equation}
H \psi (q) = E \psi (q)\, ,
\end{equation}
together with the boundary conditions (\ref{domH}). Normalized eigenstates
and corresponding eigenvalues are then given by
\begin{align}\label{PTstate}
\psi_n (q)  &= \sqrt{\frac{2}{L}} \sin \left(n\pi \frac{q}{L}\right)\, , \quad    0 \leqslant q \leqslant L \, ,\\
H \psi_n  &= E_n \psi_n \,  ,   \qquad n = 1, 2, \dotsc , 
\end{align}
with
\begin{equation}
E_n = \frac{\hbar^2\pi^2}{2mL^2} n^2 \; \equiv \; \hbar \omega n^2 \, ,  \qquad
  \omega = \frac{\hbarÊ\pi^2}{2mL^2} \equiv \frac{2\pi}{T_r} \, ,
\end{equation}
where $T_r$ is the ``revival'' time to be compared with the purely
classical round trip time.

\subsection{The quantum phase space context}

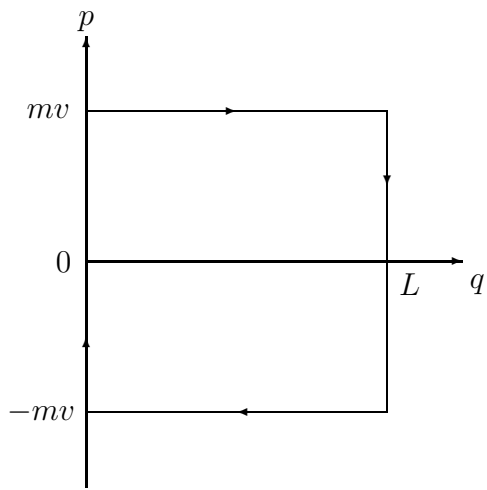
\begin{figure}

\centering

\setlength{\unitlength}{1cm}
\begin{picture}(16,8)
\put(6,3){\begin{picture}(10,6)

\put(0,0){\vector(1,0){5}}
\put(0,0){\vector(0,1){3}}
\put(0,2){\vector(1,0){2}}
\put(2,2){\line(1,0){2}}
\put(4,2){\vector(0,-1){1}}
\put(4,1){\line(0,-1){3}}
\put(4,-2){\vector(-1,0){2}}
\put(2,-2){\line(-1,0){2}}
\put(0,-3){\vector(0,1){2}}
\put(0,-1){\line(0,1){1}}

\put(-0.3,0){\makebox(0,0){$0$}}
\put(-0.5,2){\makebox(0,0){$m v$}}
\put(-0.6,-2){\makebox(0,0){$-mv$}}
\put(0,3.2){\makebox(0,0){$p$}}

\put(4.3,-0.3){\makebox(0,0){$L$}}
\put(5.2,-0.3){\makebox(0,0){$q$}}

\end{picture}}
\end{picture}
\caption{Phase trajectory of the particle in the infinite square-well.}
\label{figure1}
\end{figure}

The classical phase space of the motion of the particle is the infinite strip
$X = [0,L]\times \R = \{x= (q,p) \; | \; q\in [0,L]\, , p \in \R\}$ equipped with
the  measure: $\mu(\rmd x) = \rmd q\, \rmd p$. A phase trajectory for a given non-zero classical
energy $E_{\mathrm{class}}= \frac{1}{2}m v^2$ is represented in the figure \ref{figure1}.

Typically, we have two phases in the periodic particle motion with a given energy: one
corresponds to positive values of the momentum, $p=mv$ while the other one is for negative values, $p=-mv$.
This observation naturally leads us to introduce the Hilbert space of two-component complex-valued
functions (or spinors) square-integrable with respect to $\mu(\rmd x)$ :
\begin{equation}
\label{twohilb}
    \mathrm{L}^2_{\C^2}(X,\mu(\rmd x))
    \simeq \C^2 \otimes \mathrm{L}^2_{\C}(X,\mu(\rmd x))
    = \bigg\lbrace \Phi(x) =\bigg(\begin{matrix} \phi_+(x)\\ \phi_-(x) \end{matrix}\bigg) ,
         \ \phi_{\pm} \in \mathrm{L}^2_{\C}(X,\mu(\rmd x))\bigg\rbrace\, .
\end{equation}

We now choose our orthonormal system as formed of the following vector-valued
functions $\Phi_{n,\epsilon} (x)$, $\kappa = \pm$,
\begin{align}
\label{ortsysiw}
\nonumber \Phi_{n,+} (x) & = \bigg(\begin{matrix} \phi_{n,+}(x)\\ 0 \end{matrix}\bigg)\, , \qquad
\Phi_{n, -} (x) = \bigg(\begin{matrix} 0\\ \phi_{n,-}(x)\end{matrix}\bigg)\, , \\
\phi_{ n, \kappa}(x) & = \sqrt{c}\, \rme^{-\frac{1}{2\rho^2}(p-\kappa p_n)^2}\,
\sin \left(n\pi\frac{q}{L}\right)\, ,\qquad \kappa = \pm\,,  \ n=1,2, \dotsc,\,\end{align}
where
\begin{equation}
\label{norm}
    c=\frac{2}{\rho L \sqrt{\pi}} , \qquad
    p_n = \sqrt{2m E_n} = \frac{\hbar \pi}{L}\, n \, ,
\end{equation}
and  the half-width $\rho > 0$ is a parameter which has the dimension of a momentum,
say $\rho = \hbar \pi \vartheta/L$ with $\vartheta >0$ a dimensionless parameter. This parameter can
be arbitrarily small (like for the classical limit) and, of course, arbitrarily large (for a very narrow well,
for instance).

The functions $\Phi_{n,\kappa} (x)$ are continuous, vanish at the boundaries $q=0$ and $q=L$ of the
phase space, and obey the essential finiteness condition (\ref{factor}):
\begin{align}
\nonumber
  0  < \mathcal{ N} (x) & \equiv \mathcal{ N} (q,p)  \equiv  \mathcal{ N}_+ (x)  + \mathcal{ N}_- (x)
        = \sum_{\kappa= \pm}\sum_{n=1}^{\infty} \Phi_{n, \kappa}^{\dag} (x) \Phi_{n, \kappa} (x) \\
    &=  c\, \sum_{n=1}^{\infty}\left[  \rme^{-\frac{1}{\rho^2}(p - p_n)^2}  +  \rme^{-\frac{1}{\rho^2}
    (p + p_n)^2}\right] \sin^2\left( n\pi\frac{q}{L}\right)<
\infty \, .
\label{factor1}
\end{align}
 The expression of $\mathcal{N}$ simplifies to :
\begin{equation}
\label{norma}
 \mathcal{N}(q,p)
    = c\; \mathcal{S}(q,p)
    = c \; \Re\left\{\frac{1}{2}\sum_{n=-\infty}^\infty \big[1 -  \rme^{\rmi  2\pi n \frac{q}{L}}\big]
    \rme^{-\frac{1}{\rho^2}(p-p_n)^2}\right\}.
\end{equation}
 It then becomes apparent that $\mathcal{N}$ and $\mathcal{S}$ can be expressed in terms of elliptic theta
 functions. Function $\mathcal{S}$  has no physical dimension whereas $\mathcal{N}$  has the same
 dimension as $c$, that is the inverse of an action.

We are now in measure of defining our vector CS's \cite{aeg}.
We  set up  a one-to-one correspondence between the functions $\Phi_{n,\kappa}$'s and two-component states
\begin{equation}
\label{twocomp}
| n, \pm \rangle \deq |\pm\rangle \otimes | n \rangle\, , \qquad |+\rangle = {\,1\, \choose 0}\, , \quad |-\rangle = {\,0\, \choose 1}\, ,
\end{equation}
forming an orthonormal basis of some separable Hilbert space of the form $\mathcal{K}= \C^2 \otimes \mathcal{H}$. The
latter  can be viewed also as the  subspace of $\mathrm{L}^2_{\C^2}(X,\mu(\rmd x))$ equal to the closure of the linear
span of the set of $\Phi_{n,\kappa}$'s. We choose the following set of $2\times 2$ diagonal real matrices for our
construction of vectorial CS's:
\begin{equation}
\label{opF}
\mathrm{F}_n(x)
 = \begin{pmatrix}
    \phi_{n, +} (q,p)   &  0 \\
    0   &   \phi_{n, -}(q,p)  \end{pmatrix} \, .
\end{equation}
Note that $\mathcal{N}(x) = \sum_{n=1}^{\infty} \mathrm{tr}(\mathrm{F}_n(x)^2)$.
Vector CS's, $| x , \chi \rangle \in \C^2 \otimes \mathcal{H}= \mathcal{K} $, are
now defined for each $x \in X$ and $\chi \in \C^2$ by the relation
\begin{equation}
 | x , \chi \rangle = \frac{1}{\sqrt{\mathcal{N}(x)}} \; \sum_{n=1}^{\infty}
          \mathrm{F}_n (x)\; |\chi\rangle \otimes |n\rangle \; .
\label{def-vcs}
\end{equation}
In particular, we single out the two orthogonal CS's
\begin{equation}
\label{vectcs}
|x, \kappa\rangle  =  \frac{1}{\sqrt{\mathcal{N}(x)}} \sum_{n=1}^{\infty} \mathrm{F}_n(x) | n, \kappa\rangle \, , \qquad \kappa = \pm\, .
\end{equation}
By construction, these states also satisfy  the infinite square well boundary conditions,
namely $|x, \kappa \rangle_{q=0}=|x,\kappa\rangle_{q= L}=0$.
Furthermore they fulfill the normalizations
\begin{equation}
\label{normVCS}
 \langle x,\kappa| x,\kappa \rangle = \frac{\mathcal{N}_{\kappa}(x)}{\mathcal{N}(x)}\, ,
 \qquad \sum_{\kappa = \pm } \langle x,\kappa| x,\kappa \rangle = 1\, ,
\end{equation}
and the resolution of the identity in $\mathcal{K}$:
\begin{align}
\nonumber \int_X | x\rangle\langle x| \mathcal{N}(x) \mu(\rmd x)
 &=\sum_{\kappa,\kappa' = \pm } \sum_{n,n'=1}^\infty \int_{-\infty}^\infty\int_0^L \mathrm{F}_n(q,p)
 \mathrm{F}_{n'}(q,p) | n, \kappa\rangle\langle n', \kappa'| \rmd q \rmd p \\
 &= \sum_{\kappa = \pm }\sum_{n=1}^\infty | n, \kappa\rangle\langle n, \kappa|
 = \sigma_0 \otimes \mathbb{I}_\mathcal{H}= \mathbb{I}_\mathcal{K}\, .
\end{align}
where $\sigma_0$ denotes the $2\times 2$ identity matrix consistently with the Pauli matrix notation
$\sigma_{\mu}$ to be used in the following.

\subsection{Quantization of classical observables}

The quantization of a generic function $f(q,p)$ on the phase space is given by the expression (\ref{oper}),
that is for our particular CS choice:
\begin{align}
\label{VCSquant}
\nonumber &\widehat{f}
  = \sum_{\kappa = \pm }\int_{-\infty}^\infty \int_0^L \,    f(q,p)| x,\kappa\rangle\langle x, \kappa | \mathcal{N}(q,p) \rmd q \rmd p
  \\
& = \sum_{n,n'=1}^{\infty} | n \rangle\langle n' | \otimes \begin{pmatrix}
 \widehat{f}_+ &  0  \\
   0   & \widehat{f}_- \end{pmatrix}  \, ,
\end{align}
where
\begin{equation}
 \widehat{f}_{\pm}=\int_{-\infty}^\infty \rmd p \int_0^L \rmd q \, \phi_{n,\pm}(q,p) f(q,p) \overline{\phi_{n',\pm}}(q,p)\, .
\end{equation}
For  the particular case in which $f$ is function of $p$ only,  $f(p)$, the operator is given by
\begin{align}
\nonumber \widehat{f}
 & = \sum_{\kappa = \pm }\int_{-\infty}^\infty \int_0^L \,  f(p)| x,\kappa\rangle\langle x, \kappa | \mathcal{N}(q,p) \rmd q \rmd p
 \nonumber\\
&= \frac{1}{\rho\sqrt{\pi}}\,  \sum_{n=1}^{\infty} | n \rangle\langle n | \otimes
\begin{pmatrix} \widehat{f}_+ & 0 \cr
 0 & \widehat{f}_- \end{pmatrix}\, ,
\end{align}
with
\begin{equation}
\widehat{f}_\pm= \int_{-\infty}^{\infty} \rmd p\,   f(p) \exp\big(-\frac{1}{\rho^2}(p\mp p_n)^2\big)\,   .
\end{equation}
Note that this operator is diagonal on the $|n,\kappa \rangle$ basis.

\subsubsection{Momentum and Energy}

In particular, using $f(p)=p$, one gets the operator
\begin{equation}\label{p}
    \widehat{p}= \sum_{n=1}^{\infty} p_n \, \sigma_3\otimes| n\rangle\langle n| \, ,
\end{equation}
where $\sigma_3=\bigl(\begin{smallmatrix}  1    &  0  \\0   &  -1 \end{smallmatrix}\bigr)$ is a Pauli matrix.

For $f(p)=p^2$, which is proportional to the Hamiltonian, the
quantum counterpart reads as
\begin{equation}
\label{qkinene}
\widehat{p^2} = \frac{\rho^2}{2}\mathbb{I}_{\mathcal{K}} + \sum_{n=1}^{\infty} p_n^2 \,\sigma_0\, \otimes | n\rangle\langle n|
    = \frac{\rho^2}{2}\mathbb{I}_{\mathcal{K}} + (\widehat{p}\,)^2 \, .
\end{equation}
Note that this implies that the operator for the square of momentum does not coincide with the square of the momentum operator. Actually they coincide up to O$(\hbar^2)$.

\subsubsection{Position}

  For a general function of the position $f(q)$ our quantization procedure yields the following operator:

\begin{equation}
 \widehat{f}
    = \sum_{n,n'=1}^{\infty}  \rme^{-\frac{1}{4\rho^2}(p_n-p_{n'})^2}\left[d_{n-n'}(f)- d_{n+n'}(f)\right] \sigma_0\, \otimes | n\rangle\langle n'| \, ,
\end{equation}
where
\begin{equation}
d_{m}(f)\equiv \frac{1}{L}\int_{0}^{L} f(q)\cos\Big(m\pi\frac{q}{L}\Big) \rmd q\, .
\end{equation}In particular, for $f(q)=q$ we get the ``position'' operator
\begin{equation}
\widehat{q}
    = \frac{L}{2}\mathbb{I}_{\mathcal{K}} - \frac{2 L}{\pi^2} \sum_{n, n' \geq 1, \above0pt n+n'=2k+1}^{\infty}
    \rme^{-\frac{1}{4\rho^2}(p_n-p_{n'})^2} \left[\frac{1}{(n-n')^2}-\frac{1}{(n+n')^2}\right]\,  \sigma_0\, \otimes | n\rangle\langle n'| \, ,
\end{equation}
with $k\in \mathbb{N}$. Note the appearance of the classical mean value for the position on the diagonal.

\subsubsection{Commutation rules}

\begin{figure}

 \includegraphics[scale=.7,angle=90]{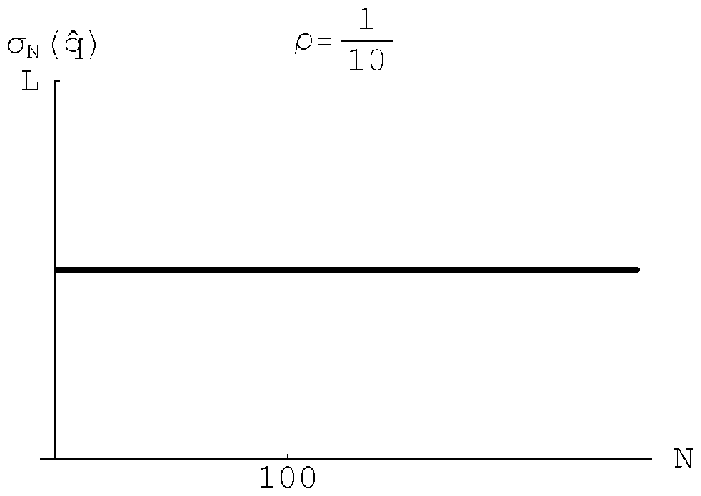}\qquad
 \includegraphics[scale=.7,angle=90]{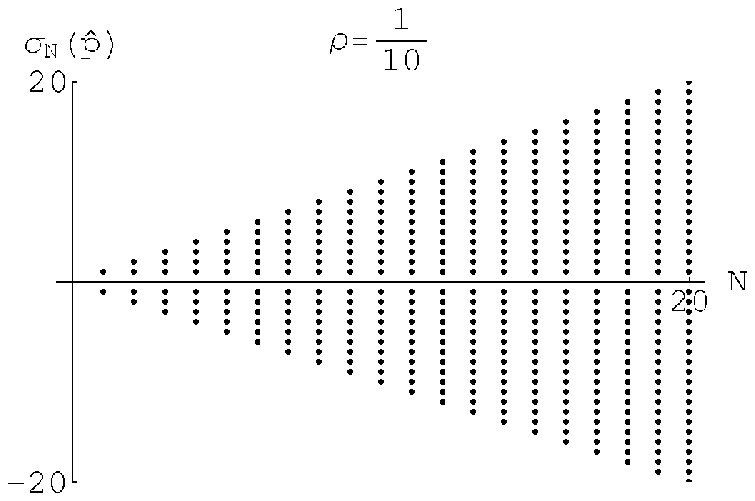}\qquad
 \includegraphics[scale=.7,angle=90]{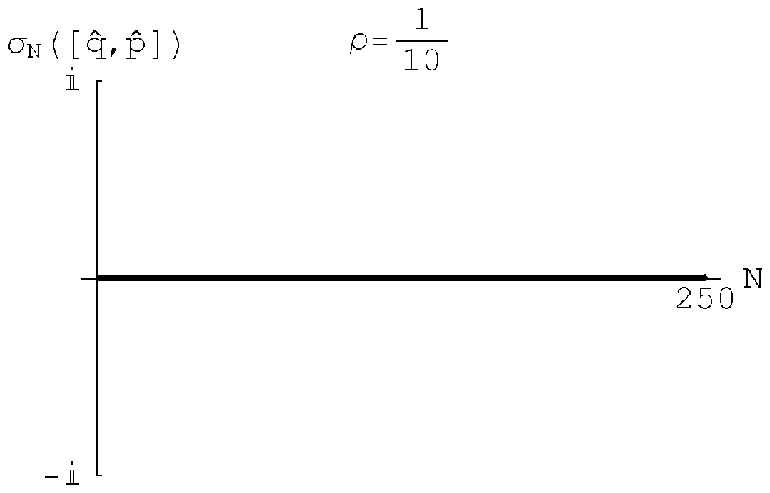}

 \includegraphics[scale=.7,angle=90]{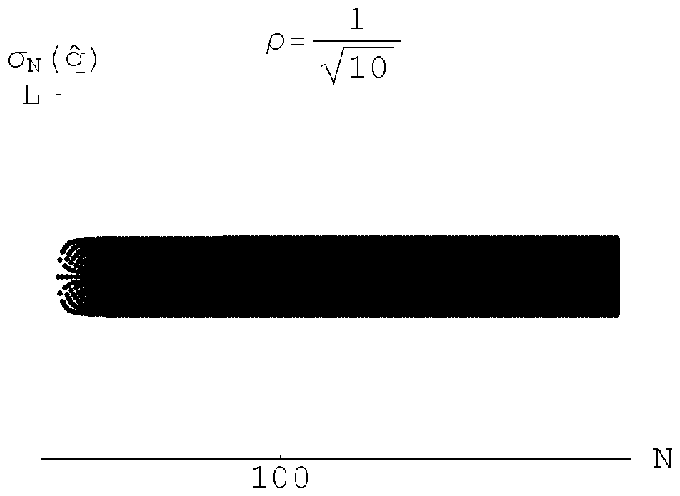}\qquad
 \includegraphics[scale=.7,angle=90]{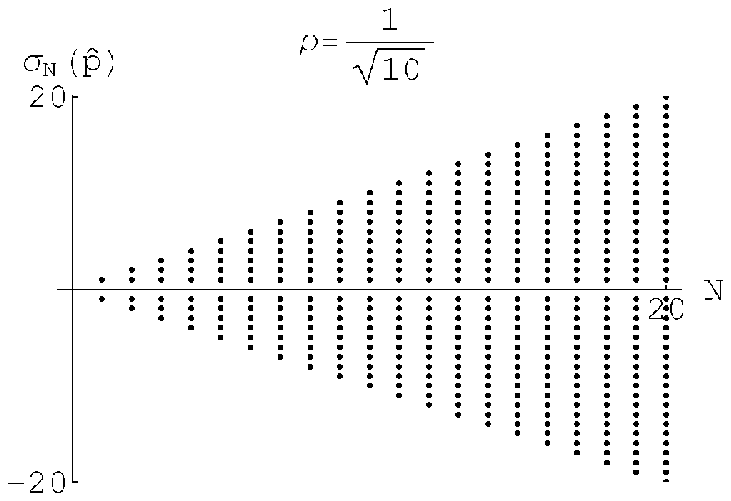}\qquad
 \includegraphics[scale=.7,angle=90]{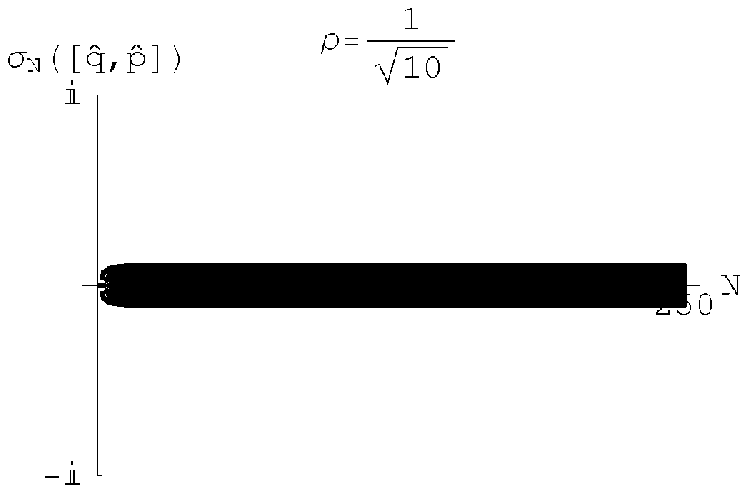}

    \caption{\label{EV1}
    Eigenvalues of $\widehat{q}$, $\widehat{p}$ and $[\, \widehat{q},\widehat{p}\, ]$ for increasing values of
    the characteristic momentum  $\rho =\hbar \pi \vartheta/L$ of the system, and computed for $N\times N$
    approximation matrices. Units have been chosen such that $\hbar = 1$, $L=\pi$ so that  $\rho=\vartheta$ and $p_n=n$.
    Note that for $\widehat{q}$ with $\rho$ small, the eigenvalues adjust to the
    classical mean value $L/2$. The spectrum of $\widehat{p}$ is independent of $\rho$ as is shown in (\ref{p}). For
    the commutator, the values are purely imaginary. }
 \end{figure}

\begin{figure}
   \includegraphics[scale=.7,angle=90]{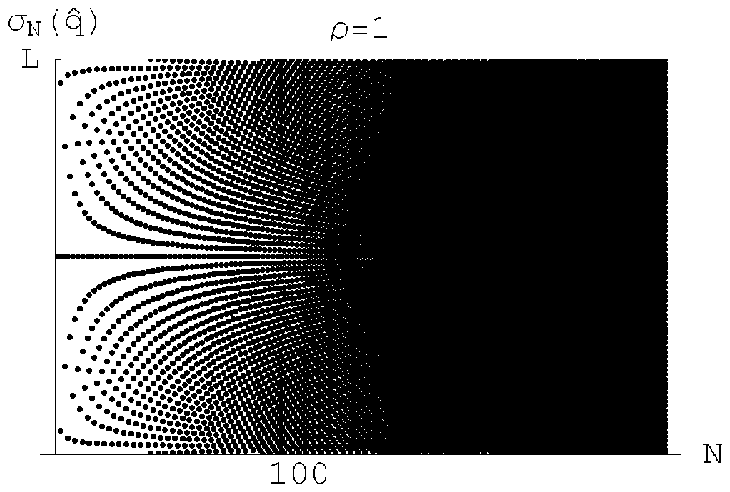}\qquad
    \includegraphics[scale=.7,angle=90]{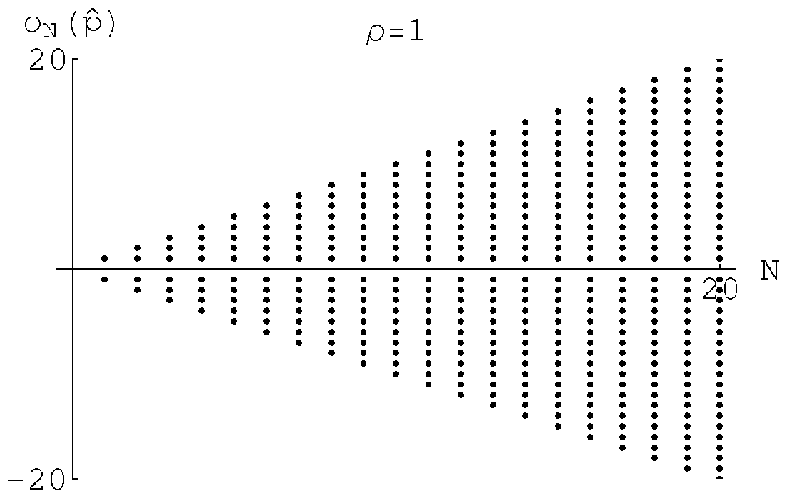}\qquad
    \includegraphics[scale=.7,angle=90]{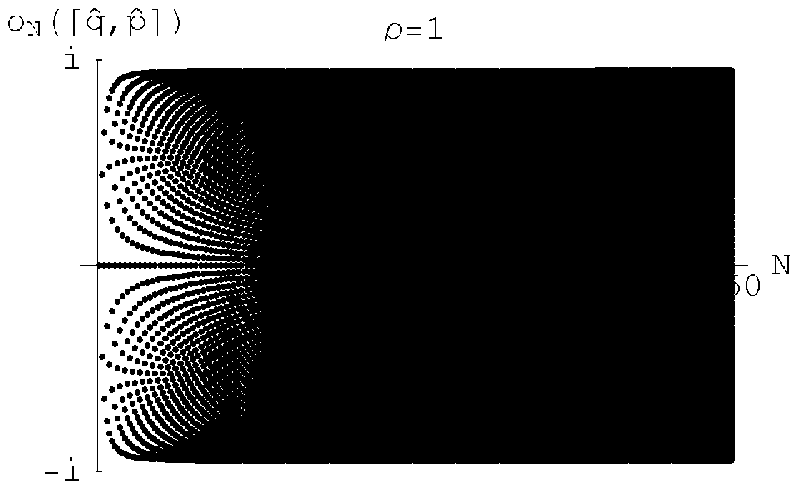}

   \includegraphics[scale=.7,angle=90]{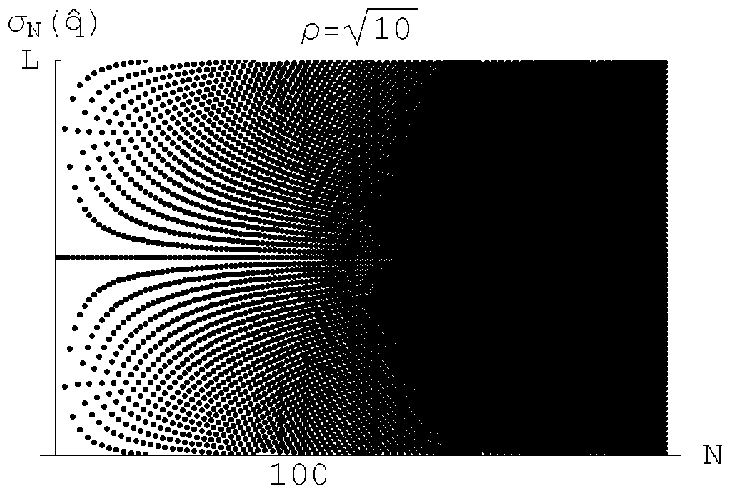}\qquad
    \includegraphics[scale=.7,angle=90]{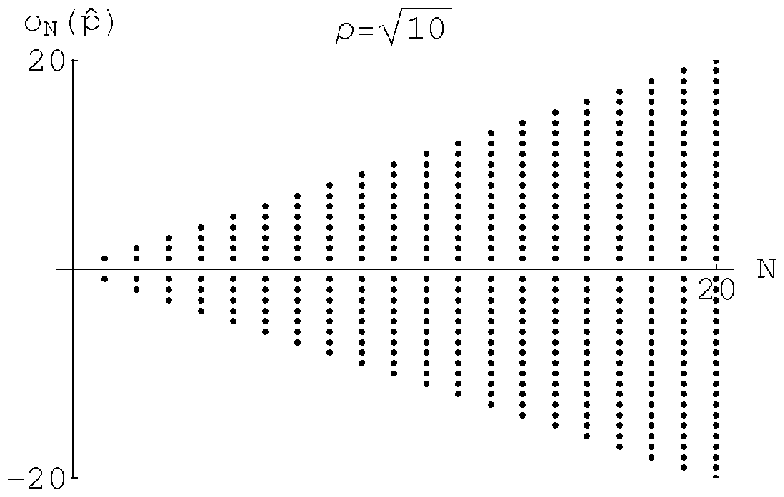}\qquad
    \includegraphics[scale=.7,angle=90]{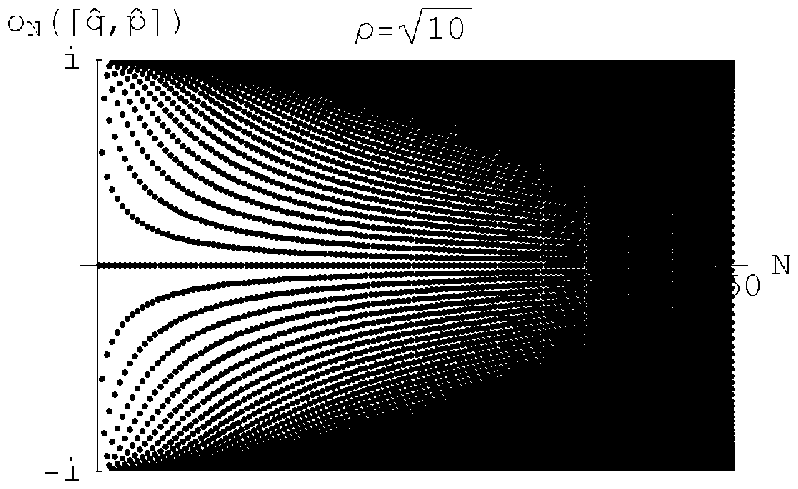}

    \includegraphics[scale=.7,angle=90]{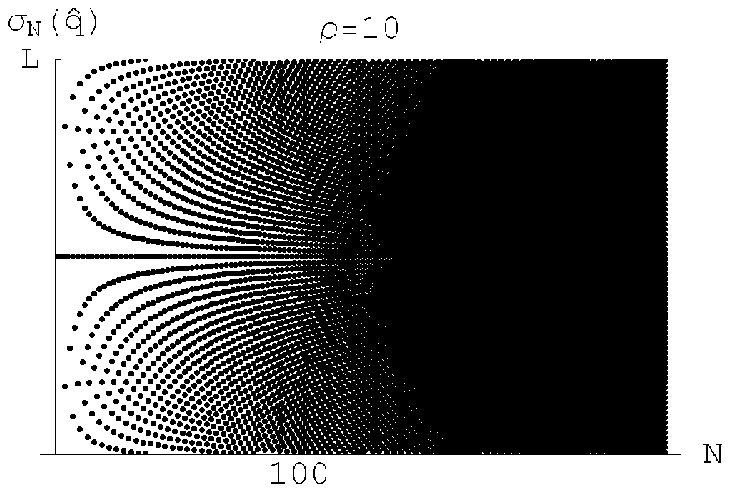}\qquad
    \includegraphics[scale=.7,angle=90]{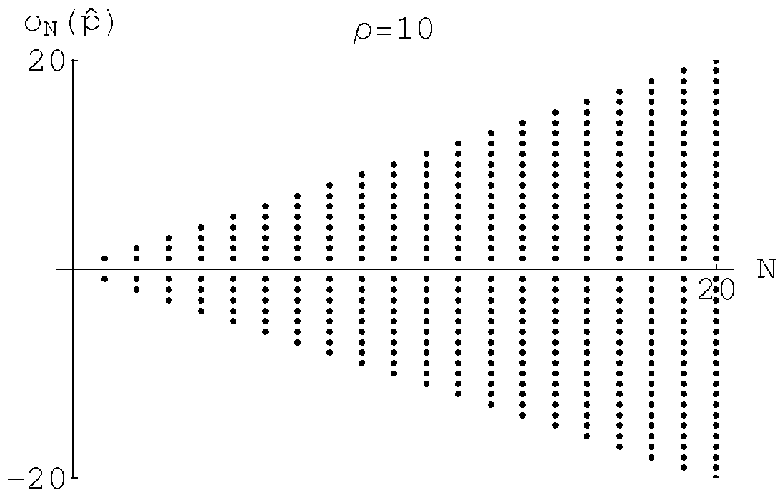}\qquad
    \includegraphics[scale=.7,angle=90]{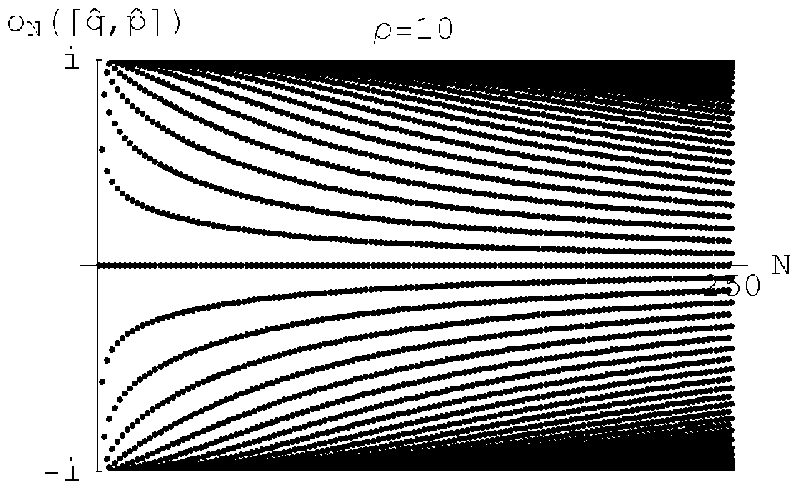}

   \caption{\label{EV2}
   Continued from figure \ref{EV1}: $N\times N$ approximation matrices eigenvalues  of $\widehat{q}$,
   $\widehat{p}$ and $[\, \widehat{q},\widehat{p}\, ]$ for increasingly larger values of
   $\rho = \hbar \pi \vartheta/L= \vartheta$ in units  $\hbar = 1$, $L=\pi$.  The spectrum of $\widehat{p}$
   is independent of $\rho$ as is shown in (\ref{p}). For the commutator, the eigenvalues  are
    purely imaginary and tend to accumulate around $\rmi\hbar$ and $-\rmi\hbar$ as $\rho$ increases.}
 \end{figure}

Now, in order to see to what extent these momentum and position operators differ from their classical (canonical)
counterparts, let us consider  their commutator:
\begin{align}
[\, \widehat{q},\widehat{p}\, ] &= \frac{2\hbar}{\pi} \!\!\!\! \sum_{n\neq n'\above0pt n+n'=2k+1}^{\infty} \!\!\!\! C_{n,n'}\, \sigma_3 \otimes | n\rangle\langle n'| \\
     C_{n,n'} &= \rme^{-\frac{1}{4\rho^2}(p_n-p_{n'})^2}(n-n')\left[\frac{1}{(n-n')^2}-\frac{1}{(n+n')^2}\right]  .
\end{align}
This is an infinite antisymmetric real matrix. The respective spectra of finite matrix approximations of this operator and of
position and momentum operators are compared in figures \ref{EV1} and \ref{EV2} for various values of the regulator $\rho=\hbar \pi \vartheta/L= \vartheta$ in units
$\hbar = 1$, $L=\pi$. When $\rho$ takes large values, one can see that the eigenvalues of $[\, \widehat{q},\widehat{p}\, ]$
accumulate around $\pm \rmi$, i.e. they become almost canonical. Conversely, when $\rho\to 0$ all eigenvalues become null, which corresponds to the classical limit.

\subsubsection{Evolution operator}
 The Hamiltonian of a spinless particle trapped inside the well is simply  $H = p^2/2m$. Its quantum counterpart
 therefore is $\widehat{H} = \widehat{p^2}/2m$. The unitary evolution operator, as usual, is given by
\begin{equation}
 U(t)
  =  \rme^{-\frac{\rmi }{\hbar} \widehat{H}t}
  =  \rme^{-i \omega_{\vth} t}\sum_{n=1}^\infty \rme^{- \frac{\rmi  p_n^2 t}{2m\hbar}}  \sigma_0\otimes | n\rangle\langle n| \, .
\end{equation}
 Note the appearance of the global time-dependent phase factor with frequency $\omega_{\vth}$ which can be compared with
 the revival frequency
 \begin{equation}
\label{globfreq}
\omega_{\vth} = \frac{\hbar \pi^2 \vth^2}{4mL^2} = \frac{\omega \vth^2}{2}\, .
\end{equation}

\section{Quantum behaviour through lower symbols}
Lower symbols are computed with normalized CS's. The latter are denoted as follows
\begin{equation}
\label{normVCS}
|x\rangle = |x, +\rangle +  |x, -\rangle\, .
\end{equation}
Hence, the lower symbol of a quantum observable $A$ should be computed as
$$\check{A}(x) = \langle x| A | x\rangle \equiv \check{A}_{++}(x) + \check{A}_{+-}(x) + \check{A}_{-+}(x) + \check{A}_{--}(x)\, .$$
This gives the following results for the observables previously considered :

\subsubsection{Position}
In the same way, the mean value of the position operator in a vector CS $|x\rangle$  is given by:
\begin{equation}
 \langle x| \widehat{q}\, | x\rangle = \frac{L}{2} - Q(q,p)\, ,
\end{equation}
where we can distinguish the classical mean value for the position corrected by the function
\begin{align}
Q(q,p)&= \frac{2L}{\pi^2}\frac{1}{\mathcal{S}} \sum_{\substack{n,n'=1, n\neq n'\\ n+n' =2k+1}}^\infty
     \rme^{-\frac{1}{4\rho^2}(p_n-p_{n'})^2}\left[\frac{1}{(n-n')^2}-\frac{1}{(n+n')^2}\right]\times\nonumber\\
     & \times \Big[  \rme^{-\frac{1}{2\rho^2}[(p-p_n)^2 + (p-p_{n'})^2]} +   \nonumber\\
     & + \rme^{-\frac{1}{2\rho^2}[(p+p_n)^2 +
     (p+p_{n'})^2]}\Big]\sin\Big( n\pi\frac{q}{L}\Big) \sin\Big( n'\pi\frac{q}{L}\Big)\, .
\end{align}
This function depends on the parameter $\vartheta$ as we show in figure \ref{VMQ} with a numerical approximation
using finite matrices. As for $\widehat p$, we calculate the dispersion defined as
\begin{equation}
\Delta Q=\sqrt{\check{q^2}-\check{q}^2}.
\end{equation}
Its behaviour for different values of $\vartheta$ is shown in figure \ref{DQ}.

\subsubsection{Time evolution of position}
    The change through time of the position operator is given by the transformation
    $\widehat{q}\,(t) := U^\dag(t)\,\widehat{q}\,  U(t)$, and differs
    from $\widehat{q}$ by the insertion of an oscillating term in the series. Its lower symbol is given by
\begin{equation}
 \langle x| \widehat{q}\,(t) | x\rangle
  = \frac{\displaystyle{L}}{2} - Q(q,p,t)\, ,
\end{equation}
where this time the series have the form
\begin{align}
Q(q,p,t) &= \frac{2L}{\pi^2}\frac{1}{\mathcal{S}} \sum_{\substack{n,n'=1, n\neq n'\\ n+n' =2k+1}}^\infty
     \rme^{- \frac{\rmi }{2m\hbar} (p_n^2-p_{n'}^2)\,  t} \,   \rme^{-\frac{1}{4\rho^2}(p_n-p_{n'})^2}\times\nonumber\\
    &\times \left[\frac{1}{(n-n')^2}-\frac{1}{(n+n')^2}\right] \sin \left(n\pi\frac{q}{L}\right) \sin \left(n'\pi\frac{q}{L}\right)\times\nonumber\\
     &\times \left[  \rme^{ -\frac{1}{2\rho^2}[(p-p_n)^2 + (p-p_{n'})^2]} + \rme^{-\frac{1}{2\rho^2}[(p+p_n)^2 +
    (p+p_{n'})^2]}\right]\, .
\end{align}
Note that the time dependence manifests itself in the form of a  Fourier series of with frequencies
$ (n^2-{n'}^2)\,\hbar \pi^2/2mL^2$. This corresponds to the circulation of the wave packet inside the well.

\subsubsection{Momentum}
The mean value of the momentum operator in a vector CS $|x\rangle$  is given by the affine combination:
\begin{align}
\label{mvmom}
 \langle x| \widehat{p}\, | x\rangle & = \frac{\mathcal{ M} (x)}{\mathcal{ N} (x)}\, , \nonumber\\
 \mathcal{ M} (x) & = c\,\sum_{n=1}^{\infty}p_n \, \left[  \rme^{-\frac{1}{\rho^2}(p - p_n)^2}  -  \rme^{-\frac{1}{\rho^2}(p + p_n)^2}\right] \sin^2\left( n\pi\frac{q}{L}\right)\, .
\end{align}
This function reproduce the profile of the function $p$, as can be seen in the figure \ref{VMP}.
We calculate then the dispersion $\Delta P$, defined as
\begin{equation}
\Delta P=\sqrt{\check{p^2}-\check{p}^2},
\end{equation}
using the mean values in a CS $|x\rangle$. Its behaviour as a function of $x$ is shown in figure \ref{DP}.

\subsubsection{Position-momentum commutator}
The  mean value of the commutator in a normalized state $\Psi = \bigl( \begin{smallmatrix}
\phi_+\\ \phi_-
\end{smallmatrix} \bigr)$ is the pure imaginary expression:
\begin{align}
\label{meanvaluecom}
\nonumber \langle \Psi | [\, \widehat{q},\widehat{p}\, ] | \Psi &\rangle  =  \frac{2\rmi \hbar}{\pi} \sum_{n\neq n'\above0pt n+n'=2k+1}^{\infty}
\rme^{-\frac{1}{4\rho^2}(p_n-p_{n'})^2}(n-n')\, \times \\
\times & \left[\frac{1}{(n-n')^2}-\frac{1}{(n+n')^2}\right]\Im{\left(\langle \phi_+|n\rangle \langle n'| \phi_+\rangle - \langle \phi_-|n\rangle \langle n'| \phi_-\rangle\right)}\, .
\end{align}
Given the symmetry and the real-valuedness of states (\ref{vectcs}), the mean value of the
commutator when $\Psi$ is one of our CS's vanish, even if the operator does not. This result is due to the symmetric
spectrum of the commutator around $0$.  As is shown in Part c) of figures \ref{EV1}, the eigenvalues of the commutator
tend to $\pm \rmi\hbar$ as $\rho$, i.e. $\vartheta$, increases. Still, there are some points with modulus less than
 $\hbar$. This leads to dispersions $\Delta Q\Delta P$ in CS's $|x\rangle$ that are no longer bounded from
below by $\hbar/2$. Actually, the lower bound of this product, for a region in the phase space as large as we wish,
decreases as $\vartheta$ diminish. A numerical approximation is shown in figure \ref{DQDP}.

\section{Discussion}

\begin{figure}
    \includegraphics[scale=.8]{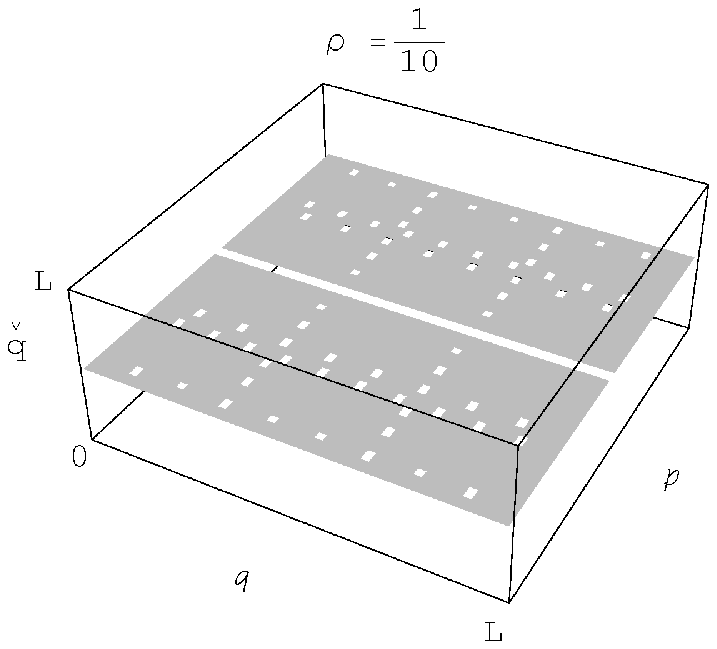}
    \includegraphics[scale=.8]{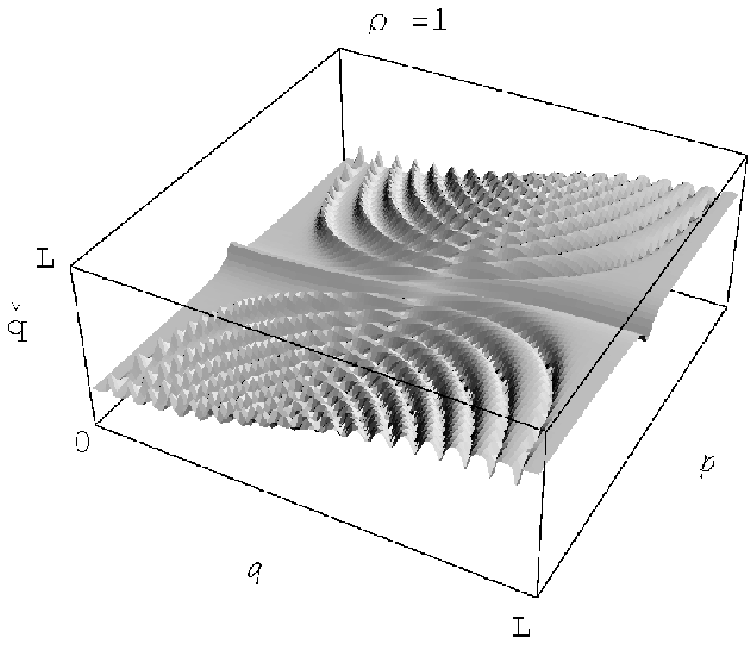}
    \includegraphics[scale=.8]{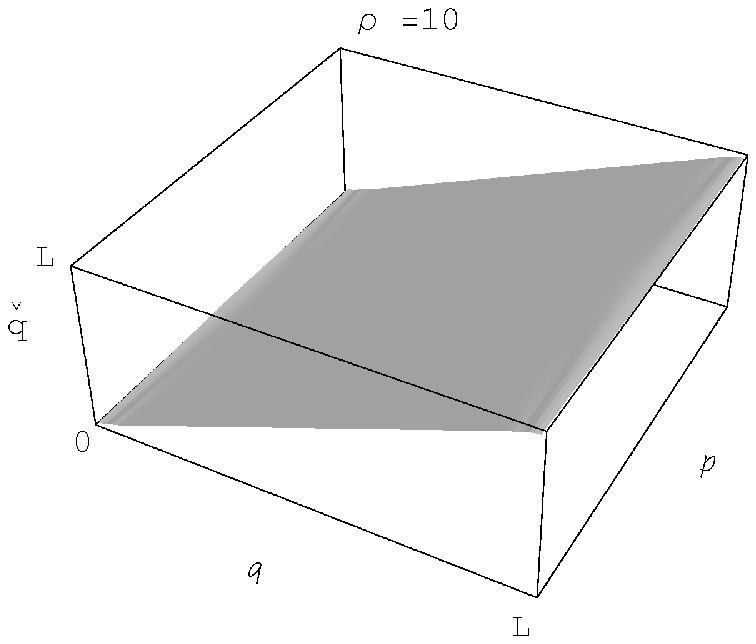}
    \caption{\label{VMQ}
    The lower symbol $\check{q}$ depicted for various values of $\rho =\hbar \pi \vartheta/L= \vartheta$
    in units  $\hbar = 1$, $L=\pi$.  Note the way the mean value fits the function $q$ when $\rho$ is large,
     and approaches the classical average in  the well for low values of the parameter.}
\end{figure}

\begin{figure}
    \includegraphics[scale=.8]{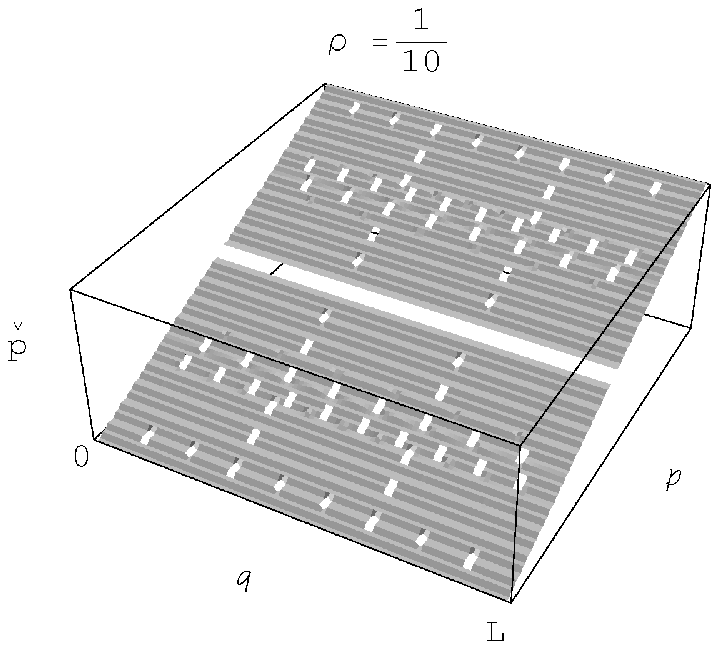}
    \includegraphics[scale=.8]{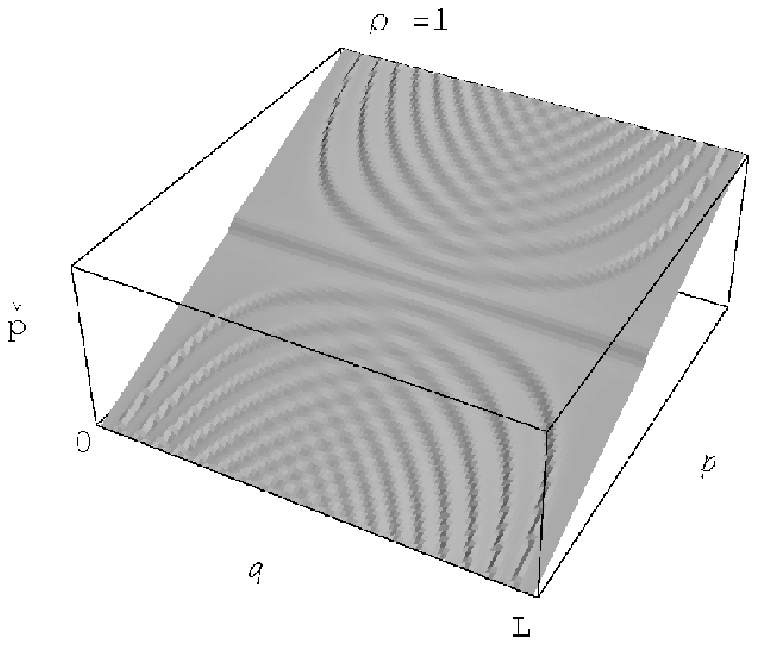}
    \includegraphics[scale=.8]{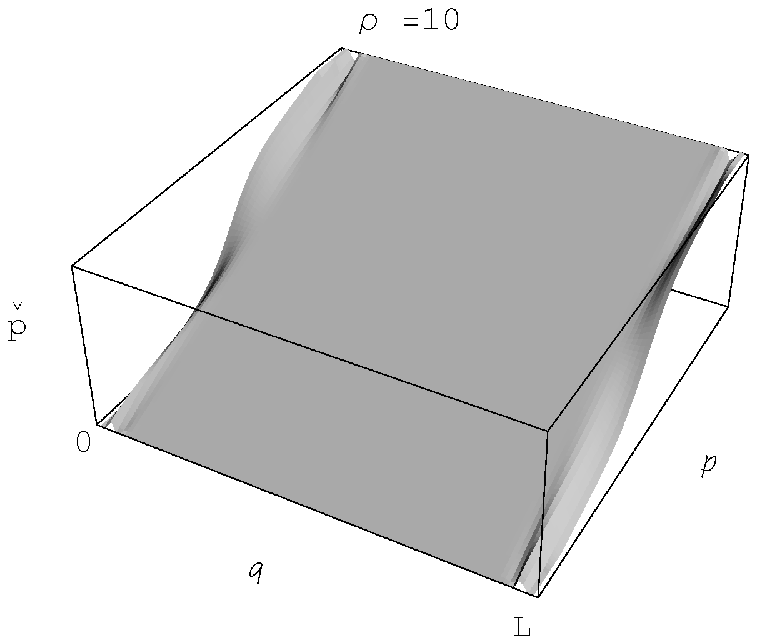}
    \caption{\label{VMP}
    The lower symbol $\check{p}$ depicted for various values of $\rho =\hbar \pi \vartheta/L =\vartheta$
    in units  $\hbar = 1$, $L=\pi$. The function becomes smoother when $\rho$ is large.}
\end{figure}

\begin{figure}
    \includegraphics[scale=.8]{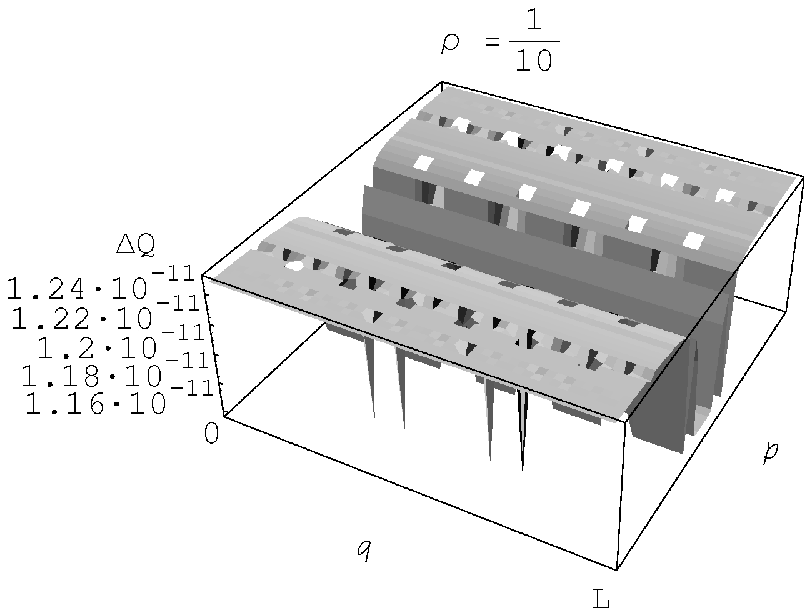}
    \includegraphics[scale=.8]{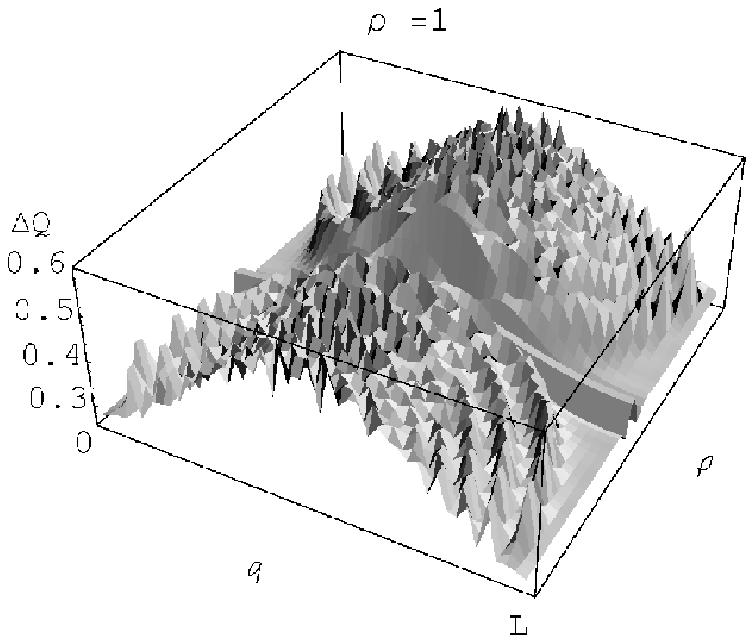}
    \includegraphics[scale=.8]{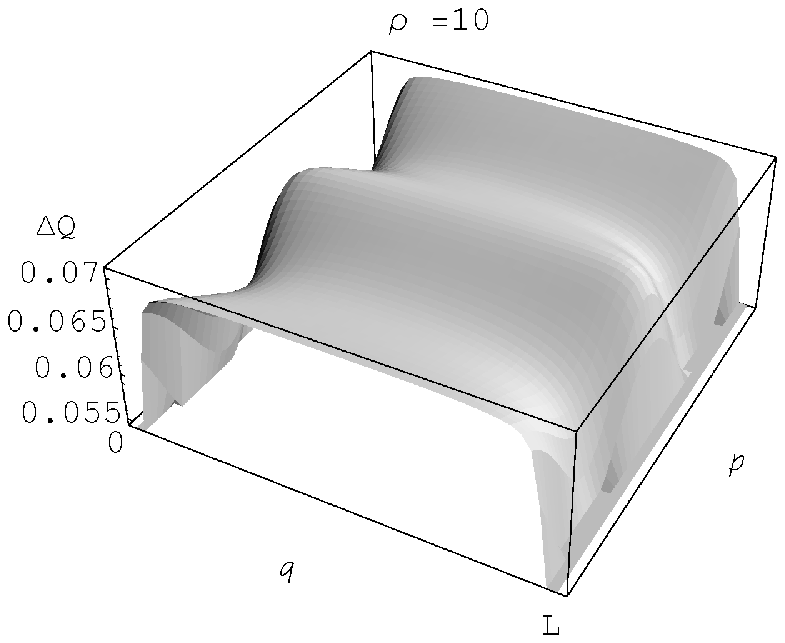}
    \caption{\label{DQ}
    Variance of $q$ depicted for various values of $\rho =\hbar \pi \vartheta/L =\vartheta$ in units
    $\hbar = 1$, $L=\pi$.
    Note how different dispersions are revealed just by changing
    the width of  the Gaussian function of the $p$ variable. Low dispersion, close to classical, is found
    for $\vartheta$
    near  $0$ and the quantum behaviour is recovered at large values of the parameter.}
\end{figure}

\begin{figure}
    \includegraphics[scale=.8]{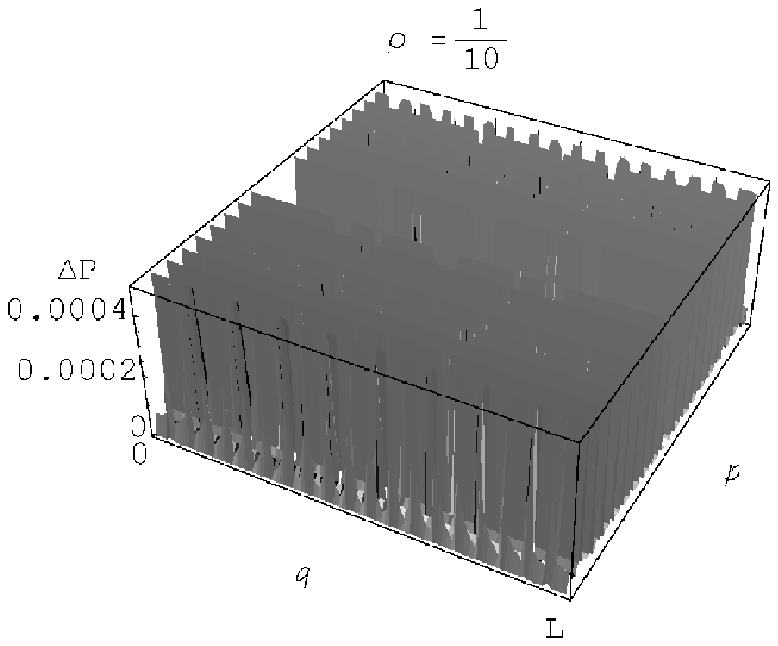}
    \includegraphics[scale=.8]{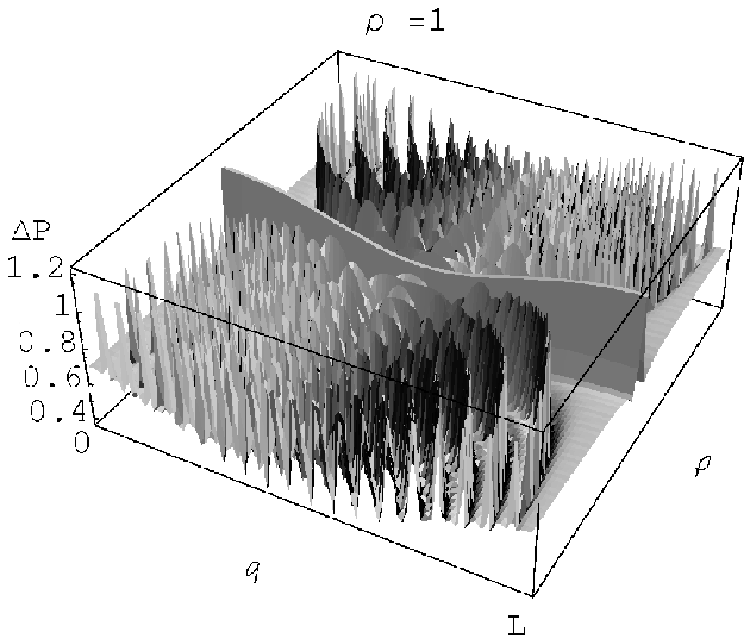}
    \includegraphics[scale=.8]{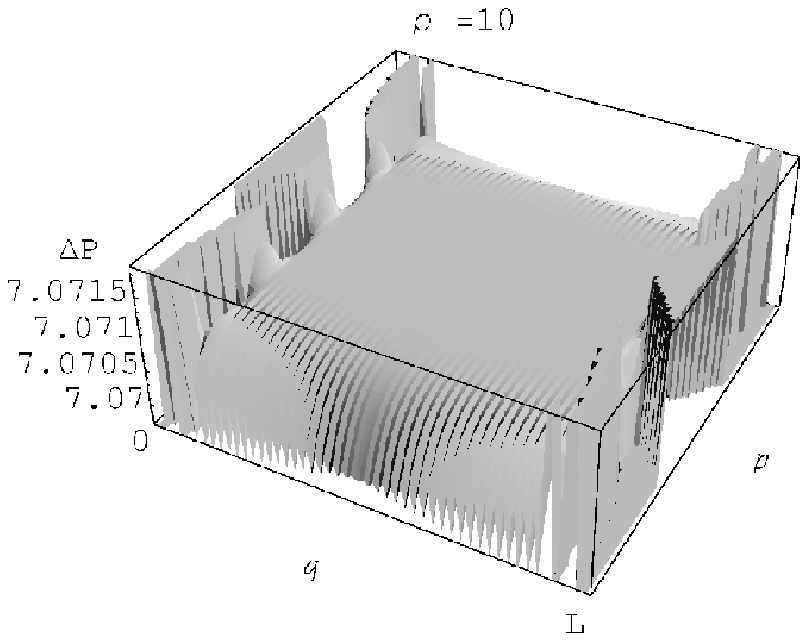}
    \caption{\label{DP}
    Variance of $p$ depicted for various values of $\rho =\hbar \pi \vartheta/L =\vartheta$ in units
    $\hbar = 1$, $L=\pi$. Consistently with $\check q$, a well localized momentum is
    found for low values of the parameter. This is actually expected  since the Gaussian becomes  very narrow. }
\end{figure}

\begin{figure}
    \includegraphics[scale=.8]{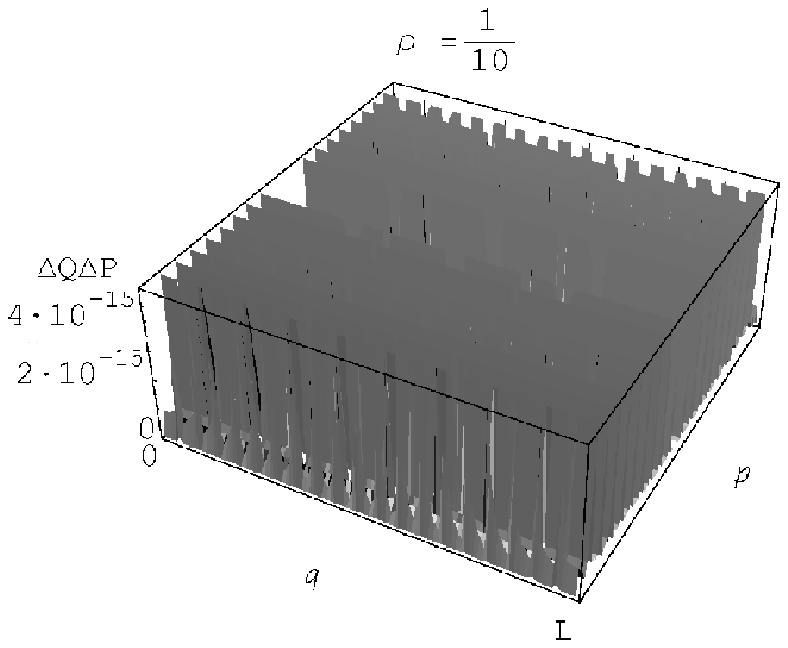}
    \includegraphics[scale=.8]{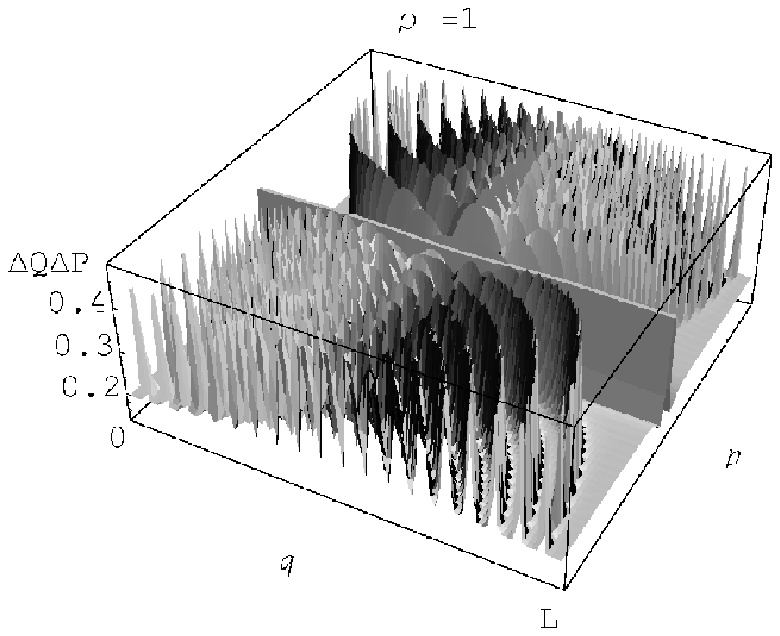}
    \includegraphics[scale=.8]{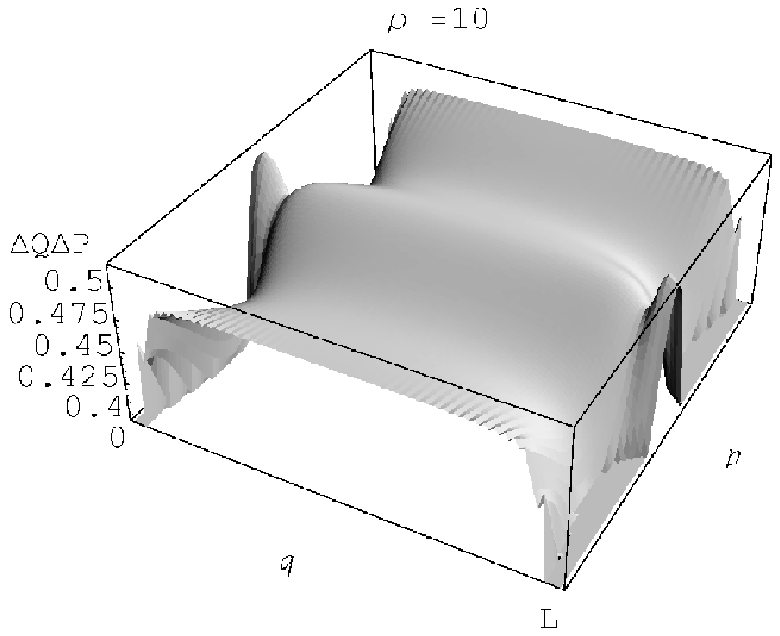}
    \caption{\label{DQDP}
    Product $\Delta Q \Delta P$  for various values of $\rho =\hbar \pi \vartheta/L= \vartheta$ in units
    $\hbar = 1$, $L=\pi$. Note the modification of the vertical scale
    from one picture to another. Again, the pair position-momentum tends to decorrelate at low values of the
    parameter, like they should do in the classical limit. On the other hand it approaches the usual quantum-conjugate
    pair at high values of $\rho$.}
\end{figure}

From the mean values of the operators obtained here, we verify that our CS quantization
gives well-behaved momentum and position operators. The classical limit is reached once the appropriate
limit for the parameter $\vartheta$ is found. If we consider the behaviour of the observables as a function of
the dimensionless quantity  $\vartheta = \rho L/\hbar\pi$, at the limit $\vartheta \to 0$ and when the Gaussian
functions for the momentum become very narrow, the lower symbol of
 the position operator is  $\check{q} \sim L/2$. This corresponds to the classical average value position in the well.
 On the other hand, at the limit
 $\vartheta \to \infty$, for which  the involved Gaussians spread to  constant functions, the mean value
 $\langle x|\hat{q}|x\rangle$ converges numerically  to the function $q$. In other words, our position operator yields
 a fair quantitative description   for the quantum localization within  the well. The lower
 symbol $\langle x|\hat{p}|x\rangle$ behaves as a stair-step function for $\rho$ close to $0$ and progressively fits
 the function $p$ when $\rho$ increases. These behaviours are well illustrated in the figures \ref{VMQ}
 and \ref{VMP}. The effect of the parameter $\vartheta$ is also noticeable in the dispersions of $\widehat q$ and
 $\widehat p$. Here, the variations  of the full width at half maximum of the Gaussian function reveal different dispersions for the operators.
 Clearly, if a classical behaviour is sought, the values of  $\vartheta$ have to be chosen near  $0$. This gives localized
 values for the observables.  The numerical explorations shown  in figures \ref{DQ} and \ref{DP} give a good account
 of this modulation.
 Consistently with the previous results, the behaviour of the product $\Delta Q \Delta P$
 at  low values of $\vartheta$ shows  uncorrelated observables at any point in  the phase space, whereas
 at large values of this parameter the product is constant and almost equal to the canonical quantum lowest limit
 $\hbar/2$. This is shown in figure \ref{DQDP}.

It is interesting to note that if we replace the Gaussian distribution, used here for the  $p$ variable in the construction of the
CS's, by any positive even probability distribution $\R\in p \mapsto \varpi (p)$ such that $\sum_n  \varpi (p-n)
< \infty$ the results are not so different! The momentum spectrum is still $\Z$ and the energy spectrum has the
form $\{n^2 + \mathrm{constant}\}$. In this regard, an interesting approach combining  mathematical statistics
concepts and group theoretical constructions of CS's has been recently developed by Heller and Wang \cite{heller1,heller2}.

The work presented here has possible applications to those particular physical problems where the square well is
used as a model for impenetrable barriers \cite{bryant}, in the spirit of what has been done in \cite{thilo}.

The generalization to higher-dimensional infinite potential wells is more or less tractable, depending on the
geometry of the barriers. This includes quantum dots and other quantum traps. Nevertheless, we believe that the
simplicity and the universality of the method proposed in the present work should reveal itself useful for this purpose.

Author Garc\'{\i}a de Le\'on wishes to acknowledge the Consejo Nacional de
Ciencia y Tecnolog\'{\i}a (CONACyT) for its support.

\pagebreak

\appendix


\end{document}